\documentclass{INTERSPEECH2023}

\interspeechcameraready 
\usepackage{algorithm}
\usepackage{algorithmic}
\usepackage{multirow}
\usepackage{cite}

\title{Pseudo-Siamese Network based Timbre-reserved Black-box Adversarial Attack in Speaker Identification}
\name{Qing Wang, Jixun Yao, Ziqian Wang, Pengcheng Guo, Lei Xie$^*$\thanks{* Lei Xie is the corresponding author.}}
\address{
  Audio, Speech and Language Processing Group (ASLP@NPU), School of Computer Science, \\
Northwestern Polytechnical University (NPU), Xi'an, China}
\email{qingwang@nwpu-aslp.org, \{yaojx, zq\_wang\}@mail.nwpu.edu.cn, \\guopengcheng1220@gmail.com, lxie@nwpu.edu.cn}

\begin{document}

\maketitle
 
\begin{abstract}
%qualified attack->machine and human black-box is more practical, so in this study, we 
% Speaker identification (SID) has shown promising performance in recent years, but with these advancements come new challenges such as the threat of different kinds of attacks.
%, such as spoofing attacks and adversarial attacks.  
In this study, we propose a timbre-reserved adversarial attack approach for speaker identification (SID) to not only exploit the weakness of the SID model but also preserve the timbre of the target speaker in a black-box attack setting.
Particularly, we generate timbre-reserved fake audio by adding an adversarial constraint during the training of the voice conversion model. Then, we leverage a pseudo-Siamese network architecture to learn from the black-box SID model constraining both intrinsic similarity and structural similarity simultaneously. The intrinsic similarity loss is to learn an intrinsic invariance, while the structural similarity loss is to ensure that the substitute SID model shares a similar decision boundary to the fixed black-box SID model. The substitute model can be used as a proxy to generate timbre-reserved fake audio for attacking. 
%The loss function used to train the substitute speaker classifier includes both intrinsic and structural similarity losses to ensure that the classifier is similar to the fixed black-box speaker classifier. The intrinsic similarity loss is computed by comparing the predictions of the substitute and black-box speaker classifiers on the same speech samples, while the structural similarity loss encourages agreement on the dependent feature shared between the two classifiers. 
%based timbre-reserved adversarial attack for the black-box setting. We leverage a pseudo-Siamese network architecture to learn from the black-box
%constrain both structural similarity and intrinsic similarity at the same time.
%Additionally, our approach operates in a black-box setting, where no knowledge of the target system's architecture or parameters is assumed, making it more practical in real-world scenarios. We demonstrate the effectiveness of our approach through extensive experiments on publicly available datasets, achieving high success rates in attacking state-of-the-art speaker identification systems while maintaining a low perceptual distance between the original and perturbed signals.
%In particular, 
%to conduct an adversarial attack via the pseudo siamese network in speaker identification
Experimental results on the Audio Deepfake Detection (ADD) challenge dataset indicate that the attack success rate of our proposed approach yields up to 60.58\% and 55.38\% in the white-box and black-box scenarios, respectively, and can deceive both human beings and machines.
%the proposed timbre-reserved adversarial attack

%exploit pseudo network to learn a substitute model from the black-box SID model,
%In particular, we 

\end{abstract}

\section{Introduction}
Speaker identification (SID)~\cite{reynolds1995robust, hansen2015speaker} is a process of determining the identity of the person who spoke a particular speech. 
As a type of biometric identification, it is critical to ensure the security of the speaker identification system. SID system usually confronts various kinds of attacks, such as spoofing attacks~\cite {wu2012detecting, wu2015spoofing, wu2017asvspoof} and adversarial attacks~\cite{szegedy2013intriguing, goodfellow2014explaining}.
%Speaker identification is a vital component of many security and authentication systems, including biometric authentication, access control, and forensic investigation. 
%spoofing attack
Spoofing attack commonly includes impersonation, replay, voice conversion, and speech synthesis.
%adversarial attack
Recently, adversarial attack has emerged as a significant threat to the accuracy and reliability of speaker identification systems, which can be defined as malicious attempts to deceive a machine learning model, including speaker identification systems, by exploiting their vulnerabilities.

Many researchers have successfully conducted adversarial attacks on SID systems~\cite{kreuk2018fooling, wang2019adversarial, abdullah2019practical, li2020universal, xie2020real, li2020adversarial, jati2021adversarial}. 
Das \textit{et al.}~\cite{das2020attacker} gave an overview of various types of attack on speaker verification focusing on potential threats of adversarial attacks and spoofing countermeasures from the attacker's perspective.
%In~\cite{li2020universal}, Li \textit{et al.} explored the existence of the universal adversarial perturbations (UAPs) in speaker recognition systems and proposed to generate UAPs by learning the mapping from the low-dimensional normal distribution to the UAP subspace via a generative model.
%Xie \textit{et al.}~\cite{xie2020real} proposed the real-time, universal, and robust adversarial attack against DNN-based speaker recognition system by adding audio-agnostic universal perturbations. 
%made the DNN-based speaker recognition system can identify the speaker as any target label by adding audio-agnostic universal perturbations on speakers' voice input. 
Li \textit{et al.}~\cite{li2020practical} launched a practical and systematic adversarial attack against speaker recognition systems and integrated the estimated room impulse response into the adversarial example training for over-the-air attack. 
%a generative network to learn the mapping from the low-dimensional normal distribution to the UAPs subspace, then synthesize the UAPs to perturb any input signals to spoof the well-trained speaker recognition model with high probability.
%proposed to generate universal adversarial perturbations (UAPs) by learning the mapping from the low-dimensional normal distribution to the universal perturbation subspace via a generative model. 
%%%后续还有方法considering the human perceptibility of sound.
To constrain the perceptibility of the adversarial perturbation and perform targeted speaker attack, in~\cite{wang2020inaudible}, Wang \textit{et al.} generated inaudible adversarial perturbations based on the psychoacoustic principle of frequency masking.
The gradients via back-propagation are essential when crafting the adversarial examples in these methods. However, in the black-box settings, it's hard to generate an adversarial example by estimating the gradient.
Zhang \textit{et al.}~\cite{zhang2022imperceptible} performed black-box waveform-level targeted adversarial attacks against speaker recognition systems by generating imperceptible adversarial perturbations based on auditory masking. 
%generated imperceptible adversarial perturbations based on auditory masking to conduct black-box waveform-level targeted adversarial attacks against speaker recognition systems.
%Furthermore, more practical conditions are explored, in~\cite{}, on-thr-air
In~\cite{chen2021real}, Chen \textit{et al.} proposed FAKEBOB to craft adversarial examples and conducted a comprehensive and systematic study of the adversarial attacks on speaker recognition systems to understand their security weakness in the practical black-box setting. 

Moreover, inspired by the aforementioned spoofing attack and adversarial attack, generating quality fake audio to effectively attack the SID model requires the ability to deceive both machines and humans simultaneously. %Inspired by the aforementioned attacks, creating quality fake audio to effectively attack the SID model requires the ability to deceive both machines and humans simultaneously.
To deceive machines, we need to consider the downstream SID task and ensure that the fake audio possesses a distinctive speaker attribute that makes the SID model give the desired decision.
From the human perception perspective, fake audio with noticeably different timbre or text from the target speaker's real audio is easily detectable. Thus, timbre and text information also need to be taken into account when conducting attacks on the SID model.
%A qualified attack in SID needs to deceive both machine and human beings. %In~\cite{}, Wang \textit{et al.} proposed a timbre-reserved adversarial attack in SID, which can exploit the weakness of the deep neural network and also can reserve the timbre of the target speaker. The setting
%exploit pseudo network to learn a substitute model from the black-box SID model,
%In particular, we 
%To this end, 
%In this study, we proposed an approach for adversarial attacks in speaker identification (SID) that aims to exploit the vulnerabilities of SID models while preserving the timbre of the target speaker in a black-box attack scenario. 

%%%%%%%对抗扰动加入vc训练之后，生成的fake audio听不到扰动，

To this end, in this study, we propose a novel timbre-preserved adversarial attack approach for speaker identification to exploit the weakness of the SID model while preserving the timbre of the target speaker in the black-box attack setting.
To achieve this, we generate timbre-preserved fake audio by incorporating an adversarial constraint during the training of the voice conversion (VC) model. We then utilize a pseudo-Siamese network architecture to learn from the black-box SID model, constraining both intrinsic similarity and structural similarity simultaneously. The intrinsic similarity loss is aimed at achieving an intrinsic invariance, while the structural similarity loss ensures that the substitute speaker classifier shares a similar decision boundary to the fixed black-box speaker classifier. By using the substitute speaker classifier as a proxy, we can generate timbre-preserved fake audio for the black-box classifier.
%The substitute speaker classifier is trained using a pseudo-Siamese network architecture, which learns from the black-box speaker classifier. This architecture is used to constrain both intrinsic and structural similarity between the substitute speaker classifier and the black-box classifier.
%The intrinsic similarity loss is used to ensure that the posterior probability of the same speaker remains consistent before and after any transformations, similar to the approach used in training the black-box speaker classifier. The structural similarity loss is employed to encourage agreement on the dependent feature shared between the substitute and black-box classifiers.
%By training the substitute speaker classifier in this way, it is hoped that it will be able to predict speaker labels for the adversarial examples generated by the attack process with high accuracy, even though it has not directly observed these examples during training.
%To achieve this, we propose a timbre-reserved adversarial attack using a pseudo-Siamese network architecture that learns from the black-box SID model, constraining both structural and intrinsic similarity simultaneously. Furthermore, we generate timbre-reserved fake audio by adding an adversarial constraint during the training of the VC model. 
%We conduct experiments , the 
Experimental results on a partial Aishell-3 dataset~\cite{shi2020aishell} indicate that the substitute speaker classifier is proximate to the black-box speaker classifier. Meanwhile, the results of experiments conducted on the Audio Deepfake Detection (ADD) challenge dataset~\cite{yi2022add} show that the fake audio generated based on our proposed method, which preserves the original timbre, can successfully deceive both humans and machines with a comparable attack success rate.

%In Section \uppercase\expandafter{\romannumeral2}, related works are introduced. 
The rest of the paper is organized as follows. In Section~\ref{sec:method}, we detail the proposed timbre-reserved black-box adversarial attack in the SID system. Datasets and experimental setup are described in Section~\ref{sec:setup}. Section~\ref{sec:result} presents the experimental results and analysis. Finally, we conclude in Section~\ref{sec:conclu}.

%\section{Pseudo Siamese Network based Attack Framework}
\section{Methodology}
\label{sec:method}
\subsection{Timbre-reserved Adversarial Attack}
\label{sec:2-1}

\begin{figure}[h]
  \centering
  \includegraphics[width=7.5cm]{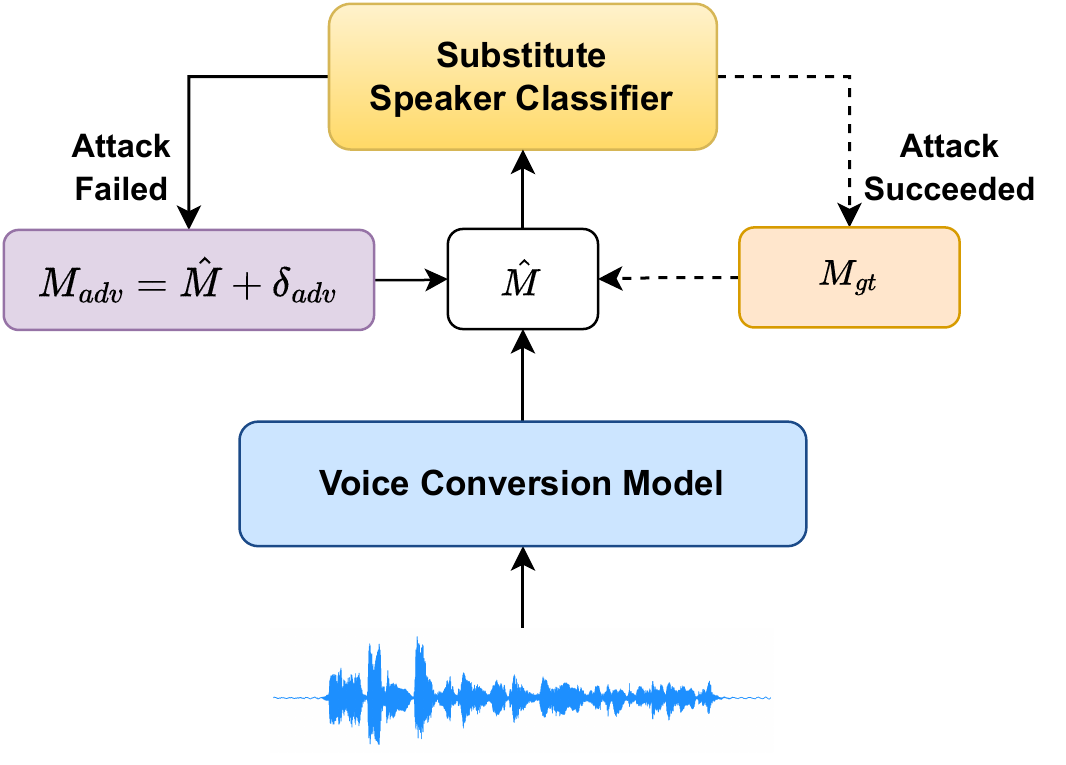}
  \caption{An overview of timbre-reserved adversarial attack.}
  \label{fig:timbre}
\end{figure}

Figure~\ref{fig:timbre} illustrates our timbre-reserved adversarial attack, which uses a non-autoregressive-based voice conversion model and an attack constraint process. We use a trained speaker identification model $f(\cdot)$ in the adversarial constraint process. Mel-spectrogram $M$ is the input, representing the VC model predicted representation, and its corresponding speaker label predicted by the SID model is $y$, while the VC target speaker label is $y'$. Adversarial constraint $\delta$ can be defined as follows:
\begin{equation}\label{commo_method}
\begin{aligned}
& \min L_{CE}(f(M+\delta), y'), \
& \text { s.t. } \quad \lVert \delta \rVert < \epsilon,
\end{aligned}
\end{equation}
here, $L_{CE}(\cdot)$ is the loss function and $\epsilon$ is the hyperparameter. For each iteration, $\delta$ is updated as follows:
\begin{equation}\label{lr}
    \delta \leftarrow \text{clip}_{\epsilon}(\delta - lr \cdot \text{sign}(\nabla_{\delta}L_{CE}(f(M+\delta), y'))).
\end{equation}
The adversarial constraint is added during the VC model training to preserve target speaker information. For joint training with the adversarial constraint, we use the Mel-spectrogram predicted by the VC model to attack the speaker classifier. If the attack fails, we add a tiny adversarial perturbation to the predicted Mel-spectrogram to generate the adversarial Mel-spectrogram $M_{\texttt{adv}}$ which can be defined as:

\begin{equation}
\begin{aligned}
& M_{\texttt{adv}}=\hat{M}+\delta_{\texttt{adv}} \\
&\text{s.t.} \quad\ \lVert  \delta_{\texttt{adv}}\  \rVert < \epsilon,
\end{aligned}
\end{equation}
here, $\delta_{\texttt{adv}}$ represents the tiny adversarial perturbation and $\epsilon$ is used to control the maximum adversarial perturbation generated.
The tiny adversarial perturbation can be optimized by: 
\begin{equation}
\begin{aligned}
\min \mathcal{L}_{\texttt{CE}}\left(f(M_{\texttt{adv}}), y^{\prime}\right) ,
\end{aligned}
\end{equation}
where $\mathcal{L}_{\texttt{CE}}$ aims to make the $M_{\texttt{adv}}$ fool the well-trained speaker identification system into predicting a specified target label. 
Therefore, the joint training with adversarial constraint can be optimized by the following loss function:

\begin{equation}\label{tot_loss}
    \mathcal{L}_{\texttt{adv}}=\left\{  
        \begin{array}{lr}  
             ||M_{gt}-\hat{M}||_{1}, \quad if\ \ succeeded, &  \\  
             ||M_{\texttt{adv}}-\hat{M}||_{1}, \quad if\ \ failed. &    
        \end{array}  \right.
\end{equation}
When $f(\hat{M}) \neq y'$, in other words, the attack failed, we add an adversarial perturbation to the Mel-spectrogram as the adversarial constraint.
%of the failed attack, 
In order to force the predicted Mel-spectrogram $\hat{M}$ of the VC model can be classified to the target speaker, we expect to minimize the L1 loss between the predicted Mel-spectrogram $\hat{M}$ and the Mel-spectrogram with the adversarial perturbation $M_{\texttt{adv}}$ so that the VC model can fool the well-trained speaker identification system.
The adversarial perturbation $\delta_{\texttt{adv}}$ is optimized by Equation~\ref{lr} until the predicted label $f(\hat{M})$ is the target label.
In contrast, when the speaker classifier gives a prediction of the target speaker label, which means the attack succeeds, the VC model is optimized only using the original reconstruction loss.

When in a more practical scenario, there is no prior knowledge of the black-box setting speaker classifier's architecture or parameters.
As a result, we train a substitute speaker classifier to conduct the black-box adversarial attack.
%the speaker classifier is in a black-box setting, where no knowledge of the target system's architecture or parameters is assumed. 

\subsection{Pseudo-Siamese Network based Substitute Speaker Classifier}

\begin{figure}[h]
  \centering
  \includegraphics[width=\linewidth]{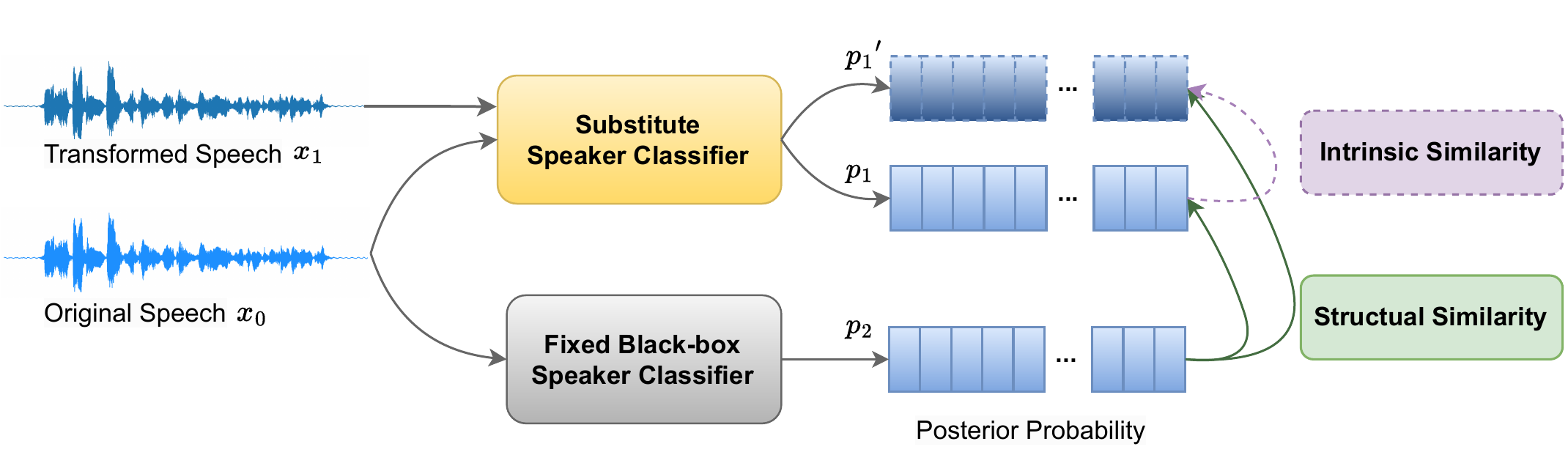}
  \caption{An overview of pseudo-Siamese network based substitute speaker classifier.}
  \label{fig:pseudo}
\end{figure}

Figure~\ref{fig:pseudo} is an overview of the pseudo-Siamese network based architecture to learn both the \textbf{intrinsic similarity} and \textbf{structural similarity} from the black-box speaker classifier. 
In particular, the intrinsic similarity enables learning of intrinsic invariance, while the structural similarity guarantee that the substitute speaker classifier shares a similar decision boundary to the fixed black-box classifier. Since there is no prior knowledge of the target SID system's architecture or parameters, we can only access the input and output of the black-box speaker classifier but not the internal processing.

The intrinsic similarity loss is minimized by comparing the original speech and transformed speech, with the aim of assisting the substitute speaker classifier in learning with disturbed labels.
It is expected that the posterior probability of the same speaker should remain consistent before and after any transformations.
Specifically, for a given speech sample $x_0$, a transformation was applied by adding Gaussian noise, resulting in a transformed speech sample $x_1$, which increases the diversity of the data distribution. 
The intrinsic similarity loss is calculated by comparing the output probabilities of $x_0$ and $x_1$:
\begin{equation}
\begin{aligned}
\mathcal{L}_{\text{ins}}=KL(p_1|| {p_1}') ,
\end{aligned}
\end{equation}
where $\mathcal{L}_{\text{ins}}$ is the intrinsic similarity loss and $p_1$ and ${p_1}'$ is the output posterior probability of $x_0$ and $x_1$ from the substitute speaker classifier, while $KL(\cdot)$ is the Kullback-Leibler (KL) divergence~\cite{moreno2003kullback}. 

The structural similarity loss is calculated by comparing the posterior probability distributions of speech samples from substitute and black-box speaker classifiers. In particular, for a given speech sample $x_0$ and transformed speech sample $x_1$, its structural similarity loss is defined as follows: 
\begin{equation}
\begin{aligned}
\mathcal{L}_{\text{aux}}=KL({p_1}'|| p_2) ,
\end{aligned}
\end{equation}

\begin{equation}
\begin{aligned}
\mathcal{L}_{\text{str}}=KL(p_1|| p_2) + \mathcal{L}_{\text{aux}} ,
\end{aligned}
\end{equation}
where $\mathcal{L}_{\text{str}}$ is the structural similarity loss, $\mathcal{L}_{\text{aux}}$ is the auxiliary similarity loss, and $p_2$ is the posterior probability of original speech $x_0$ from the black-box speaker classifier. 
The final training objective is defined as follows:
%the substitute and black-box speaker classifiers before and after transformation. Specifically, for a given speech sample, its intrinsic similarity loss is defined as the L2 distance between the output probabilities of the two classifiers before and after applying a transformation to the speech sample. This loss encourages the substitute speaker classifier to produce consistent output probabilities with the black-box classifier.
\begin{equation}
\begin{aligned}
\mathcal{L}_{\text{total}}=\mathcal{L}_{\text{ins}}+\mathcal{L}_{\text{str}} ,
\end{aligned}
\end{equation}
where $\mathcal{L}_{\text{total}}$ is the total loss to constrain both structural and intrinsic similarities.

The pseudo-Siamese network architecture is utilized because it can effectively capture the similarity between the different outputs of the same speech samples from different classifiers. %The structural similarity loss is used to constrain the substitute speaker classifier to have similar behavior as the black-box speaker classifier, while the intrinsic similarity loss is used to enhance the diversity of the data distribution.
Overall, the training procedure for the substitute speaker classifier aims to learn a model that behaves similarly to the black-box speaker classifier in terms of predicting speaker identities, while also being able to withstand adversarial attacks. After the substitute speaker classifier is trained, the attack loss is calculated as described in Section~\ref{sec:2-1}.

\subsection{Generation of Timbre-reserved Fake Audio}
\begin{figure}[h]
  \centering
  \includegraphics[width=6.5cm]{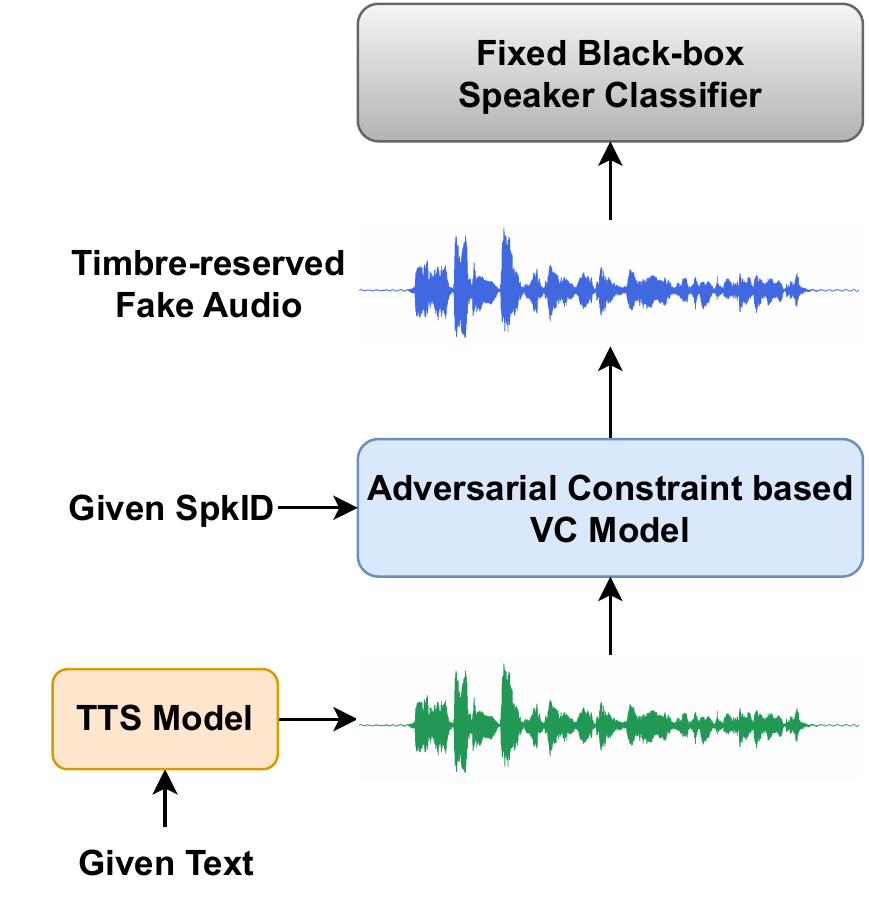}
  \caption{The generation of timbre-reserved fake audio.}
  \label{fig:generation}
\end{figure}

%The generation of timbre-reserved black-box fake audio is shown in Figure~\ref{fig:generation}. 
After the VC model with adversarial constraints is trained, the generation of timbre-reserved black-box fake audio is shown in Figure~\ref{fig:generation}. 
Firstly, given a text, we generate audio for a random speaker using a TTS system based on the FastSpeech~\cite{fs2} model. The encoder and decoder structure of the TTS model is modified to incorporate the conformer block from DelightfulTTS 2~\cite{delightfultts2}. The speaker ID and the TTS-generated audio are then used as inputs for the adversarially constrained VC model to predict the attack Mel-spectrogram.
%With the given text, we generate the audio of a random speaker by a TTS system. The TTS system is based on the FastSpeech~\cite{fs2} model and modified the encoder and decoder structure inspired by the DelightfulTTS 2~\cite{delightfultts2} conformer block. 
%Then the given speakerID and the TTS generated audio are the inputs of the adversely constrained VC model to predict the attack Mel-spectrogram. 
Finally, a HifiGAN vocoder~\cite{hifigan} is used to reconstruct the waveform from the Mel-spectrogram, resulting in timbre-reserved fake audio that can be used for the adversarial attack against the fixed black-box SID model.
%A HifiGAN vocoder~\cite{hifigan} is followed to reconstruct the waveform from Mel-spectrogram. After that, the timbre-reserved fake audios are obtained for the adversarial attack against the SID model.

%\subsection{Training Objective}

\section{Experimental Setup}
\label{sec:setup}
\subsection{Datasets}
In this study, we use AISHELL-3~\cite{shi2020aishell} to train the pseudo-Siamese network, speaker identification model, and voice conversion model. 
AISHELL-3 is a multi-speaker Mandarin Chinese audio corpus containing 88,035 recordings from 218 native speakers. 
%AISHELL-3 is used in the speaker identification model and voice conversion model training, while 
The test set of AISHELL-1~\cite{bu2017aishell} is used to evaluate the performance of the SID model.
%为了在公开数据集上建立一个攻击条件，we follow the rule of add 3.1, which is to.... 然后给定了text和speaker id，我们要用生成的数据攻击声纹模型
%For the timbre-reserved adversarial attack, the dataset of audio deepfake detection (ADD) challenge~\cite{yi2022add} is used to evaluate our proposed method. 
%Since ADD challenge corpus is an open-source dataset, and is specifically designed for deep fake audio attack and detection, we employ this corpus to establish a solution to first address the adversarial attack while remaining the timbre and also customizing the text. 
To evaluate our proposed method for the timbre-reserved adversarial attack, we employed the dataset of the audio deepfake detection (ADD) challenge~\cite{yi2022add}, which is an open-source dataset designed for deep fake audio attack and detection. 
%The training and adaptation sets for ADD Track 1 and 2 are used to train the fake detection models and the test set of these two tracks are used to evaluate the detection models.
%In Track 3.1 of ADD challenge, 10 speaker IDs and 500 texts are given to generate the fake audio. 
%In the test set of ADD challenge Track 3.1, 10 speaker IDs and 500 texts are given to generate the fake audio. We generate fake audio according to the given text and speaker identities and the fake audio can fool the fake detection model and SID model. Furthermore, the generated attack samples need also to meet the certain requirement of intelligibility and similarity.
We then evaluate our proposed method on the test set of ADD challenge Track 3.1, which consists of 10 speaker IDs and 500 texts used to generate fake audio. Our generated attack samples are required to meet certain standards of intelligibility and similarity while still being able to fool the SID model. 
By using ADD challenge dataset, we develop a solution to address adversarial attacks while maintaining timbre and allowing for customized texts.
\subsection{Setup}
The detailed experimental setup of all the models shown in Figure~\ref{fig:generation} is described as follows:
\begin{itemize}
\item TTS model: 6-layer conformer encoder-decoder structure similar to DelightfulTTS 2~\cite{delightfultts2} is used to generate waveforms from given text with a random speaker timbre.

\item VC model: Pre-trained ASR model~\cite{yao2021wenet} is used to extract speaker-independent linguistic information. A non-autoregressive VC model is adopted using an 8-layer transformer encoder-decoder structure similar to FastSpeech2~\cite{fs2} and HifiGAN vocoder with multi-band processing~\cite{hifigan}. The speaker classifier is added to adversely constrain distinctive speaker information to generation.

\item Speaker identification model: ECAPA-TDNN~\cite{desplanques2020ecapa} black-box SID model used in this study, is also used as a speaker classifier in voice conversion. EER on AISHELL-1 test set is 1.91\%. Incorporates 3 SE-Res2Block modules with channel size and bottleneck dimension set to 1024 and 256, respectively. AAM-softmax loss function~\cite{deng2019arcface} with a margin of 0.2 and scale of 30.
The substitute speaker classifier consists of six conformer~\cite{gulati2020conformer} blocks arranged in a stack followed by a statistics pooling layer which is similar to x-vector~\cite{snyder2018x}.

\item Mel-spectrogram adversarial constraint experiments: VC model learning rate in Equation~\ref{lr} is set to 8e-4 and $\delta$ is updated 1000 times for each mini-batch. $l_\infty$ norm is used to measure perturbation bound. $\epsilon$ starts from 0.8. The substitute speaker classifier is added to adversely constrain distinctive speaker information to generation.
\end{itemize}

\iffalse
\subsection{Evaluation Metrics}
%%%the larger the better
In this study, we adopt several criteria to measure the performance of various kinds of generated fake audio as follows: 

\begin{itemize}

\item Attack success rate is used to evaluate the performance of targeted attacks in speaker identification. The attack success rate is also the accuracy predicted from the SID, denoted as $Acc$. Formally, the accuracy is calculated as:
\begin{equation} 
    Acc = \frac{N_s}{N},
\end{equation}
where $N$ is the total number of fake audios we generated to test and $N_s$ refers to the number of audios attacking successfully.
The higher the $Acc$ in a targeted attack means the better the targeted attack is conducted.
\end{itemize}
\fi

\section{Experimental Result}
\label{sec:result}
\iffalse
\begin{table}[]\centering
\caption{Attack success rates (\%) of different kinds of generation methods.}
\label{tab:prediction}
\renewcommand{\tabcolsep}{0.4cm}
\renewcommand\arraystretch{1.6}
\begin{tabular}{ccc}
\toprule
Speaker classifier                                  & white\_gt\_acc & black\_white\_acc \\ \hline
Conformer SV_1                 & 77.6\%         & 0.800             \\
Conformer SV_2  & 82.1\%         & 0.842             \\\bottomrule
%Conformer SV_3      & 79.8\%         & 0.816         \\ \bottomrule  
\end{tabular}
\end{table}
\fi
\subsection{Performance of Substitute SID Model}
%We evaluate the performance of the substitute speaker classifier by comparing it with the ground truth speaker label and the black-box speaker classifier.
The performance of the substitute speaker classifier is evaluated by comparing its results to both the ground truth speaker label and the predictions made by the black-box speaker classifier, shown in Table~\ref{tab:Acc1}. 
%Table~\ref{tab:Acc1} shows the accuracy of the substitute speaker classifier prediction is evaluated by comparing it to the ground truth speaker label and the black-box speaker classifier prediction. 
We can observe that 91.44\% of the predictions of substitute and black-box classifiers are the same, while compared with the ground truth speaker label, the accuracy of the substitute classifier is 90.16\%.
This indicates the substitute SID is proximate to the black-box SID.
\begin{table}[h]
\caption{Accuracy (\%) of substitute speaker classifier prediction comparing with ground truth speaker label and black-box speaker classifier prediction.}
\vspace{-5pt}
\label{tab:Acc1}
\renewcommand{\tabcolsep}{0.4cm}
\renewcommand\arraystretch{1.1}
\begin{tabular}{lcc}
\bottomrule
            & Black-box SID  &  Ground Truth    \\ \hline
Black-box SID       & -             & 97.42 \\
Substitute SID & 91.44          & 90.16 \\\bottomrule
%Ground truth          & -             & 100   \\ \bottomrule
\end{tabular}
\vspace{-5pt}
\end{table}

\subsection{Attack Success Rate}
The attack success rate, denoted as `Acc' in Tabel~\ref{tab:Acc}, is used to measure the performance of targeted attacks in speaker identification. It represents the accuracy predicted from the SID and is calculated by dividing the number of successfully attacked audios by the total number of generated fake audios. A higher Acc indicates a better attack.

The first method used in the comparison is generated only by the vanilla VC model, which serves as the baseline for all other methods and is denoted as `VC' in Table~\ref{tab:Acc}. The second method involves the direct addition of adversarial perturbation to the fake audios generated by the VC model, using the approach proposed in~\cite{wang2020inaudible}. As the perturbation is optimized by the SID system and directly added to the waveform, it is still perceptible to human beings. This method is sub-optimal in the scenarios of deceiving both humans and machines, though represents the upper limit of adversarial attack and is denoted as `VC+adv' in Table~\ref{tab:Acc}. 
The remaining three proposed methods are based on the VC model trained with multiple types of adversarial constraints, as outlined in Section~\ref{sec:method}, are denoted as `White-box', `Black-box ($\mathcal{L}_{\text{str}}$)', and `Black-box ($\mathcal{L}_{\text{total}}$)', respectively.
As shown in Table~\ref{tab:Acc}, the attack success rate of vanilla VC model based fake audio is 29.60\%, while the VC audio with direct adversarial perturbation~\cite{wang2020inaudible} achieves 76.50\%. Meanwhile, the Acc results of our proposed timbre-reserved strategies are 60.58\%, 52.86\%, and 55.38\%, respectively. %Due to the adversarial constraint being added only during the model training instead of directly added to the attack waveform, the attack success rates of the proposed strategies have a small gap with the VC+adv method.
%%因为只是在训练模型的时候用sid模型对vc加对抗约束，而不是直接在样本上加，所以会和直接在样本上加的结果有一个gap，
%%observed that the fake audios generated through all three of our proposed timbre-reserved adversarial strategies show significant improvement compared to those generated through the original vanilla VC model, with improvements of 30.98%, 25.34%, and 36.7%, respectively.
We can observe that the fake audios generated based on all these three proposed timbre-reserved adversarial strategies are significantly improved compared to the fake audios generated by the original vanilla VC model, with improvements of 30.98\%, 23.26\%, and 25.78\%, respectively. %And our proposed methods also have comparable results with the direct adversarial perturbations to the vanilla VC generated fake audios.
Furthermore, our proposed methods yield comparable results to those obtained through direct adversarial perturbations to the fake audios generated by the vanilla VC model.

\begin{table}[]\centering
\caption{Attack success rates (\%) of different kinds of generation methods.}
\vspace{-5pt}
\label{tab:Acc}
\renewcommand{\tabcolsep}{0.45cm}
\renewcommand\arraystretch{1.5}
\begin{tabular}{ccc}
\toprule
                  & Method   & Acc (\%) $\uparrow$ \\ \hline
Baseline          & VC        & 29.60   \\
Upper limit       & VC+adv~\cite{wang2020inaudible}    & 76.50   \\ \hline
                  & White-box & \textbf{60.58}   \\
\textbf{Proposed} & Black-box ($\mathcal{L}_{\text{str}}$) &   \textbf{52.86}  \\
                  & Black-box ($\mathcal{L}_{\text{total}}$) &   \textbf{55.38} \\  \bottomrule
\end{tabular}
\vspace{-15pt}
\end{table}

\subsection{Ablation Study}
To further investigate the effectiveness of each component in the pseudo-Siamese framework, we conduct ablation studies of different components in the loss function. As shown in Table~\ref{tab:Abla}, the accuracy of the substitute speaker classifier prediction is evaluated by comparing it to the ground truth speaker label and the black-box speaker classifier prediction. 
In the ablation studies, we first remove the intrinsic similarity loss ``$\mathcal{L}_{\text{int}}$", only remaining the structural similarity ``$\mathcal{L}_{\text{str}}$", the performance degrades to 82.33\% and 85.11\%.
Then, when continue removing the auxiliary similarity loss ``$\mathcal{L}_{\text{aux}}$", the performance degrades to 67.19\% and 66.22\%.
%The results demonstrate that removing variants  get worse performance in both. 
Both removed losses are calculated between the posterior probability of transformed speech and of substitute and black-box speaker classifiers, which proves the effectiveness of our proposed loss function and data augmentation method.
%\vspace{-5pt}
\begin{table}[h]
\caption{Ablation studies of different components in the loss function.}
\vspace{-5pt}
\label{tab:Abla}
\renewcommand{\tabcolsep}{0.3cm}
\renewcommand\arraystretch{1.2}
\begin{tabular}{lcc}
\bottomrule
       Variants          & Ground Truth   & Black-box SID  \\ \hline
Proposed ($\mathcal{L}_{\text{total}}$)               & 90.16\% & 91.44\% \\
$-$ Intrinsic $\mathcal{L}_{\text{int}}$           & 82.33\% & 85.11\% \\
%$\ \ -$ 
$\quad -$Auxiliary $\mathcal{L}_{\text{aux}}$ & 67.19\% & 66.22\% \\ \bottomrule
\end{tabular}
\vspace{-15pt}
\end{table}

\subsection{Generated Audio Quality Evaluation}
\begin{figure}[t]
  \centering
  \includegraphics[width=8cm]{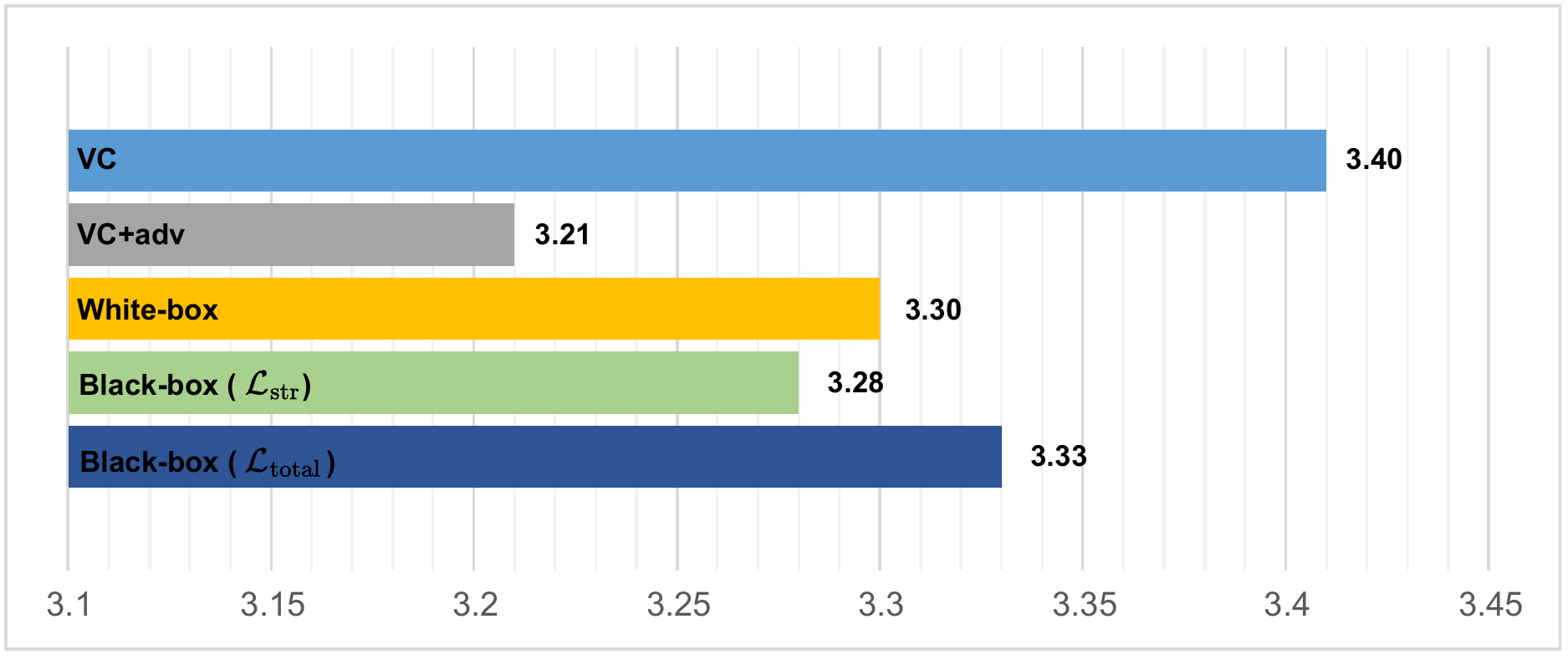}
  \caption{The o-MOS of different kinds of generation methods.}
  \label{fig:obj}
  \vspace{-15pt}
\end{figure}

We employ MOSNet~\cite{lo2019mosnet} prediction of objective Mean Opinion Score (o-MOS) to measure the quality of fake audio generated by different methods, as depicted in Figure~\ref{fig:obj}. Notably, the o-MOS results of fake audios generated through our proposed strategies are consistently superior to those generated audios with direct adversarial perturbation addition. This indicates higher quality of fake audio generated through our proposed strategies, as we integrate adversarial constraints into the training process of the VC model, thereby avoiding the direct introduction of additional human-perceptible perturbations to the fake audio.%%%黑盒和白盒是comparable的

\section{Conclusion}
\label{sec:conclu}
In this study, we propose a timbre-reserved adversarial attack on speaker identification in a black-box attack scenario. The proposed method involves using a timbre-reserved adversarial constraint during the training of the VC model to generate fake audio, along with a pseudo-Siamese network architecture to learn from the black-box SID model. The intrinsic similarity loss ensures intrinsic invariance, while the structural similarity loss ensures a similar decision boundary between the substitute speaker classifier and the fixed black-box speaker classifier. The substitute speaker classifier is used as a proxy to generate timbre-reserved fake audio for the black-box classifier. The experiments on the ADD challenge corpus show that our approach significantly improves the attack success rate compared to the vanilla VC model, without affecting the quality of the audio generated by the VC model or introducing extra noise. The proposed method offers a new perspective on adversarial attacks in speaker identification, emphasizing the importance of preserving the timbre of the target speaker during the attack.

\bibliographystyle{IEEEtran}
\bibliography{mybib}

\end{document}